# An Automated Deep Learning Approach for Bacterial Image Classification

M.TALO[1]

[1] Munzur University, Tunceli/Turkey, muhammedtalo@munzur.edu.tr

*Abstract* – Automated recognition and classification of bacteria species from microscopic images have significant importance in clinical microbiology. Bacteria classification is usually carried out manually by biologists using different shapes and morphologic characteristics of bacteria species. The manual taxonomy of bacteria types from microscopy images is time-consuming and a challenging task for even experienced biologists. In this study, an automated deep learning based classification approach has been proposed to classify bacterial images into different categories. The ResNet-50 pre-trained CNN architecture has been used to classify digital bacteria images into 33 categories. The transfer learning technique was employed to accelerate the training process of the network and improve the classification performance of the network. The proposed method achieved an average classification accuracy of 99.2%. The experimental results demonstrate that the proposed technique surpasses state-of-the-art methods in the literature and can be used for any type of bacteria classification tasks.

## I. INTRODUCTION

There are numerous bacteria species in our body and around us. Some bacteria species are beneficial for human life while others are harmful. Beneficial bacteria species help digestion of foods, fermentation of dairy products, drug production, etc. The harmful species of bacteria are the main cause of various diseases. Therefore, the classification of bacterial species is very important especially for human life.

Biologists try to identify and classify various bacteria types, which have different biochemistry and shapes. They use different attributes of bacteria for classification. For instance, the shape of the bacterial cells (spiral, cylindrical and spherical), the size and structure of the colonies formed by the bacteria are examined to differentiate bacteria species [1]. The cells of some bacteria types have different size and structure depending on environmental conditions. Some species of bacteria are very similar in shape. Although each bacteria species has its own characteristics, the biochemical reactions that bacteria carry out and the metabolic activities they perform together help to classify species [2]. However, the classification of bacteria species is not an easy task even by experienced specialists.

The classification of bacterial species with the help of computer-aided systems would provide great convenience for biologists. Accurate and rapid classification of bacterial species is of great importance, particularly early detection and treatment of diseases caused by bacteria. Different image processing approaches, such as traditional machine learning methods and modern deep learning techniques, have been used in various studies for the recognition and classification of bacterial types. For the automated detection of bacteria types, Ahmed et al. [3], employed the Fisher discriminant analysis method to extract features and a Support Vector Machine (SVM) was used for classification. Bruyne et al. [4] utilized Random Forest and SVM machine learning methods to identify three different types of bacteria, Fructobacillus, Leuconostoc, and Lactococcus. The authors reported different recognition accuracies, which vary between 94% and 98%. Goodacre et al. [5], used different spectrometry and artificial neural networks (ANN) approaches to identify bacteria types. They have reached the highest identification accuracy of 80%. Nie et al. [6] employed convolutional deep belief networks (CDBN) to learn high-level representations from digital bacterial images. Then they trained a convolutional neural network (CNN) to classify and segment bacterial images. Fiannaca et al. [7], proposed a classification approach based on k-mer representation and used deep learning technique to classify bacteria sequences.

The classical machine learning techniques are mainly used for the purposed of feature extraction. Then, the learned representations are passed through a classifier, such as an SVM or random forest, for classification tasks. However, in deep learning, the representations that obtained from raw images and the classification process take place on a single structure.

In recent years, the deep learning algorithms, especially convolutional neural networks are commonly used for various image processing and computer vision tasks such as identification, classification, and segmentation [8]. CNN architectures have many successful applications in the fields of health, safety, language translation, pathology, microbiology, etc. They are usually trained on a large amount of labeled data (supervised learning) using parallel computation methods with the help of GPUs.

In this study, deep learning based pre-trained ResNet-50 architecture was used to classify bacterial species from digital images. The transfer learning technique was used for the accurate and robust training process. Transfer learning method enabled to train the model with fewer data samples. The proposed method has an end-to-end structure and classifies bacterial images without requiring any pre-processing step.

## II. MATERIAL AND METHODS

### A. Bacteria Species Dataset

In this study, DIBaS bacteria species dataset [9], was used to classify digital bacteria images. The dataset is publicly available and composed of 33 bacterial species. Each bacteria type has about 20 images. All the bacteria images have the resolution of 2048×1532 pixels. Several images from DIBaS dataset is shown in Figure 1. The different bacteria species from the dataset and the number of them are given in Table 1.

The samples from the dataset were stained with Gramm method. All bacteria images were acquired from the Olympus CX31 microscope.

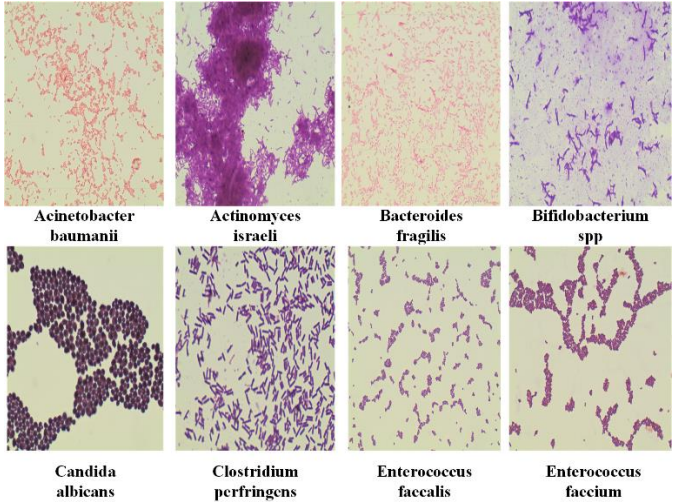

Figure 1: Sample bacteria images from DIBaS dataset

Table 1: The name of bacteria species and the number of samples from DIBaS dataset.

| Species | Number |
| --- | --- |
| Lactobacillus johnsonii | 20 |
| Listeria monocytogenes | 22 |
| Propionibacterium acnes | 23 |
| Veionella | 22 |
| Staphylococcus aureus | 20 |
| Enterococcus faecium | 20 |
| Lactobacillus gasseri | 20 |
| Streptococcus agalactiae | 20 |
| Actinomyces Israeli | 23 |
| Fusobacterium | 23 |
| Pseudomonas aeruginosa | 20 |
| Lactobacillus plantarum | 20 |
| Lactobacillus reuteri | 20 |
| Clostridium perfringens | 23 |
| Neisseria gonorrhoeae | 23 |
| Proteus | 20 |
| Acinetobacter baumanii | 20 |
| Lactobacillus casei | 20 |
| Bacteroides fragilis | 23 |
| Porfyromonas gingivalis | 23 |
| Escherichia coli | 20 |
| Lactobacillus crispatus | 20 |
| Bifidobacterium spp | 23 |
| Staphylococcus epidermidis | 20 |
| Staphylococcus saprophiticus | 20 |
| Lactobacillus salivarius | 20 |
| Lactobacillus delbrueckii | 20 |
| Lactobacillus jehnsenii | 20 |
| Candida albicans | 20 |
| Lactobacillus rhamnosus | 20 |
| Micrococcus spp | 21 |
| Lactobacillus paracasei | 20 |
| Enterococcus faecalis | 20 |

### B. Deep Transfer Learning

Deep learning is a popular approach that commonly used in image processing tasks such as detection, classification, and segmentation. Deep learning algorithms, especially Convolutional Neural Networks (CNN), automatically learn high-level feature representations form the raw data without requiring any handmade feature extractions.

A CNN is constructed by stacking a block of convolutional, polling, and fully connected layers. There are also activation functions and some other layers such as normalization and dropout, which regularize the parameters of the network for efficient training which results in high performance. The convolutional layers learn representational information from the data by applying linear convolutional operations. The convolutional operations are performed by using different filters (kernels). A feature map is created by sliding the filter through the input data. The convolutional process is a mathematical linear operation obtained by an element-wise matrix multiplication between the input image and a filter. The sum of products is calculated and the results of convolutional operations are passed through the activation functions. An activation function enables a network to learn non-linear features. For example, the rectified linear unit (ReLU) activation function is commonly placed after the convolutional layer and describe as:

$$\text{ReLU}(x) = \max(0, x)$$

for a given input x. The Softmax activation function, which attached to the final layer of convolutional neural networks, outputs a probability distribution for each target class. A pooling layer reduces the size of the activation map spatially by extracting the important features from the image. The feature map size is reduced by pooling layer. Therefore, the pooling layer also decreases the computation cost of the network during training. For instance, the average pooling layer calculates the average of numbers in a feature map. The fully connected layers, which located at the end of the network, classify the images into various categories.

Training deep learning based CNN model from starch requires a large amount of labeled data and high computational power. The network weights (parameters) at the beginning of training are randomly initialized with some small values, which are usually between 0 and 1, as a starting point, then they are updated by the help of an optimization algorithm. This training process can take several weeks depending on the training data size and available computer hardware. The transfer learning technique [10] tackles these problems. Instead of training a model from scratch, a CNN model which was previously trained for a different but related large-scale task can be used. It is not always possible to find numerous annotated data for training. The transfer learning technique enables to train networks which have a relatively small number of data.

In deep learning models, as the number of layers gets deeper, the accuracy of the model increases up to a point. Deep learning models learn the high-level features in deeper layers. However, an increase in the number of layers causes information loss after a certain point. This also complicates the training process of a network and the performance of the model may even degrade. ResNet architecture has overcome these problems. The ResNet architecture was released by He et al. [11], got the first place in ILSVRC and COCO challenges in 2015. ResNet introduced residual connections, also known as skip connection. The residual shortcuts add extra connections among residual blocks of a network for the information flow through the whole network. The skip connections were enabled to train deeper neural networks with even 1001-layers.

### III. EXPERIMENTAL SETUP

In this study, ResNet-50 pre-trained CNN architecture is employed to classify bacteria species into 33 categories using the transfer learning technique. The convolutional and the pooling layers of ResNet-50 model are transferred to the new model. The fully connected layers of ResNet-50 is removed from the end of the network and replaced with a brand new fully connected layer that outputs a 33 unit tensors using the Softmax activation function to classify bacteria species. Additionally, two dropout layers have been added to the network to avoid overfitting during training. Once the overfitting occurs, the model does not generalize on unseen data, that is, it memorizes the training samples but fails on test data. The first and the second dropout rates are set to 25% and 50%, respectively.

The dataset consists of a total of 689 bacterial images. 80% of bacteria images (552) was allocated for the training set and the rest of the images, 20% (137) used for the validation set. The learning rate hyper-parameter is randomly set to 1e-3. The learning rate hyper-parameter control the speed of the parameter update. If it is set to be too low, the model classification performance proceeds so slowly. On the contrary, if it is chosen to be too high, the model would have missed the local minimum and diverges. The RMSprop optimizer algorithm is used to update the parameters of the network during the training. The proposed ResNet-50 pre-trained model is trained and tested on Ubuntu 16.04 server using NVIDIA GeForce GTX 1080 TI graphic card. In the training and testing of the proposed ResNet-50 model, the Python programming language based PyTorch [12] libraries and Fastai [13] framework were used.

### IV. RESULTS

5-fold cross-validation technique was used to evaluate the performance of proposed pre-trained ResNet-50 model. The experiments were repeated five times and the average of five trials on validation sets was given as a classification performance for the overall model. The ResNet-50 pre-trained model was trained for 50 epochs, that is, the model examined each training image 50 times. The 5-fold validation accuracy and the training time for each fold are given in Table 2. A visualization of the validation accuracy graph for five folds and training and validation loss graph for Fold-5 are shown in Figure 2 and Figure 3, respectively.

Table 2: Training time and the classification accuracy for each fold.

| Fold | Training Time (min : sec) | Validation accuracy (%) |
|---|---|---|
| Fold-1 | 31:55 | 99.27 |
| Fold-2 | 32:17 | 99.27 |
| Fold-3 | 31:30 | 98.54 |
| Fold-4 | 31:42 | 99.27 |
| Fold-5 | 31:35 | 99.27 |
| **Average** | 31:48 | 99.12 |

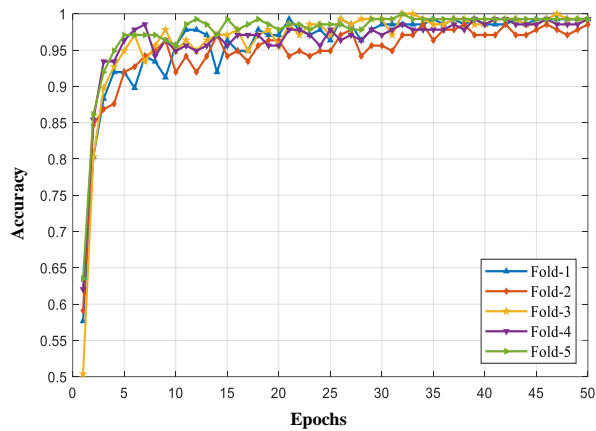

Figure 2: 5-Fold training accuracy of ResNet-50 model.

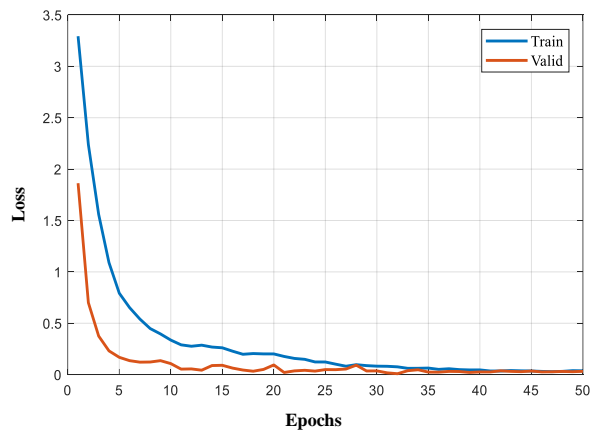

Figure 3: The training and validation loss graph for Fold-5.

As a result of all experiments, 99.12% average classification accuracy was obtained on the test sets. The average training time of ResNet-50 took about 31 minutes and 48 seconds. It can be seen from Figure-4 that, there is no overfitting during training, i.e., the training and validation losses were well-balanced and decreased during training. For the detailed performance analysis, the confusion matrix for Fold-5, which obtained using validation data, is given in Figure 4.

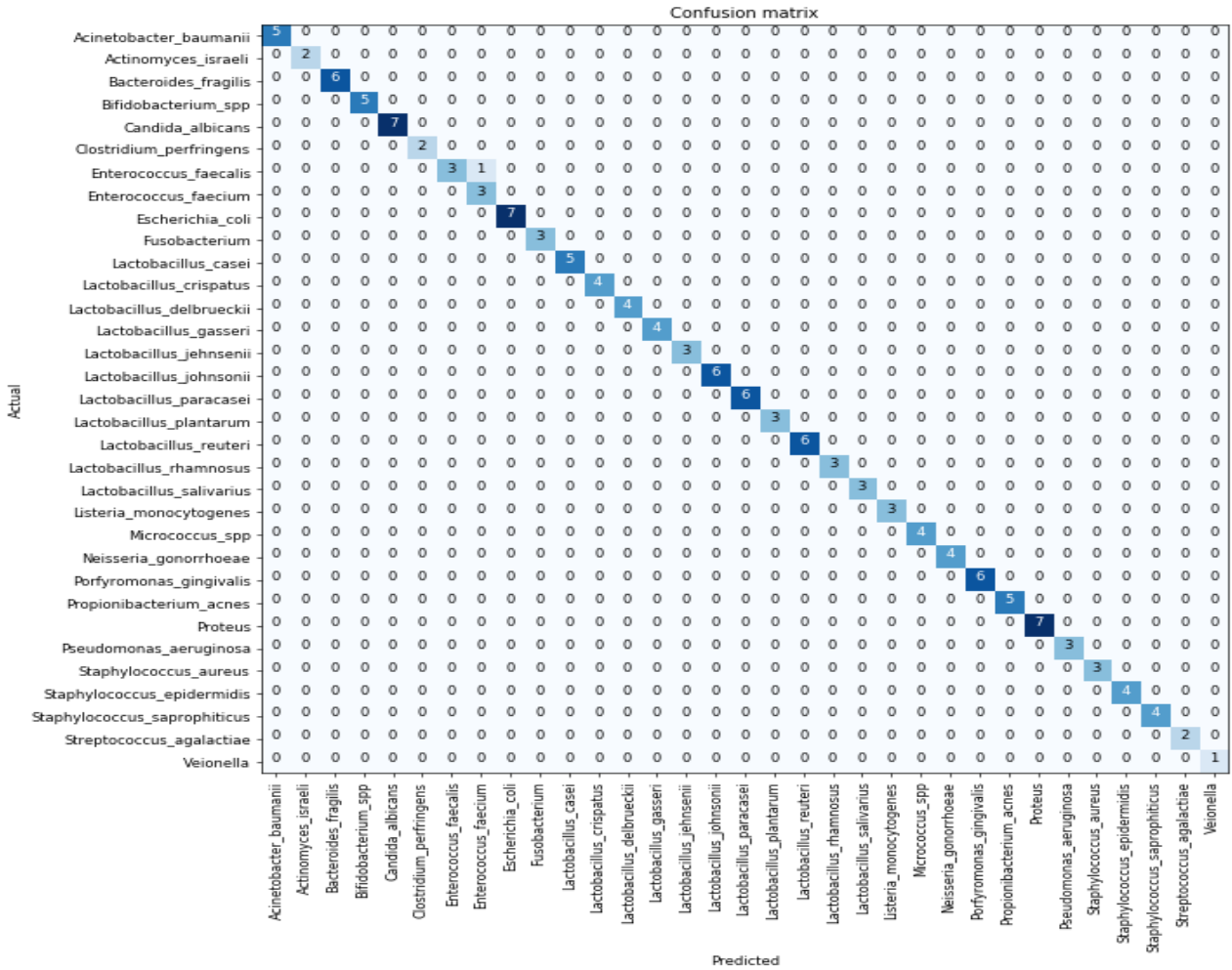

Figure 4: The obtained confusion matrix using the validation set for Fold-5.

The confusion matrix shows that, on the test set for Fold-5, all the bacteria images were classified correctly by the proposed ResNet-50 pre-trained model except one bacteria image, Enterococcus Faecalis. The model prediction for misclassified bacteria image was Enterococcus Faecium. Note that both bacteria types, Enterococcus Faecalis, and Enterococcus Faecium, comes from the enterococcus genre.

## V. DISCUSSION

The classification performance of this study was compared with other state-of-the-art studies, which classified the bacterial images into different categories using DIBaS dataset. Table-3 shows the existing studies in the literature that conducted on the same bacterial image dataset.

In 2017, Zielinski et al. [9] applied convolutional neural networks to DIBaS bacterial image dataset. The authors applied fisher vector (FV), local image descriptors and the pooling encoder to acquire image descriptors. Then the Support Vector Machines (SVM) and Random Forest (RF) methods were used to classify species of bacteria into 33 classes. They have reported 97.24± 1.07% classification accuracy.

Mohamed et al. [14], proposed histogram equalization and Bag-of-words (BoW) methods to extract features from the DIBaS image dataset. They used Support Vector Machine to classify bacterial images into 10 categories. The authors provided an average classification accuracy of 97%.

In 2018, Nasip et al. [15], used deep learning based VggNet and AlexNet pre-trained CNN architectures to classify bacterial images into 33 categories. The authors obtained the highest classification accuracy of 98.25% via VggNet.

Table 3: The studies that are conducted on DIBaS dataset.

| Study | Methods | Classes | Accuracy (%) |
|---|---|---|---|
| Zielinski et al. [9] (2017) | CNN, SVM, RF | 33 | 97.24 |
| Mohamed et al. [14] (2018) | BoW, SVM | 10 | 97 |
| Nasip et al. [15] (2018) | VggNet | 33 | 98.25 |
| **The proposed (2019)** | **Resnet-50** | **33** | **99.12** |

Table 3 demonstrates that the proposed approach has achieved an average classification accuracy of 99.12%. The performance of the pre-trained ResNet-50 model also evaluated using precision, recall, and F1-score. The average precision, recall, and F1-score value for the validation sets was 99%. As a result, the proposed approach outperforms the existing studies in the literature for the same dataset.

## VI. CONCLUSION

Bacteria are all over life. The beneficial species of bacteria have positive effects on human life. For example, they heal diseases and help digestion. However, harmful bacterial species affect human life negatively by causing diseases. The automated classification of the different type of bacterial species has great importance. In this study, ResNet-50 pre-trained CNN model is employed to classify bacteria species into 33 classes. This model has reached an average classification accuracy of 99.2%. The proposed study automatically classifies bacteria types without using any pre-processing technique. This model is ready to be examined in the field of clinical microbiology to automatically classify bacterial images.